\documentclass[]{spie}  

\usepackage{amsmath,amsfonts,amssymb}
\usepackage{graphicx}
\usepackage[colorlinks=true, allcolors=blue]{hyperref}
\usepackage{comment}
\usepackage{subcaption}
\usepackage{wrapfig}

\usepackage{url}
\usepackage{hanging}
\usepackage{lipsum}
\usepackage{multicol}
\usepackage{multirow}
\usepackage{overpic}
\usepackage[inline]{enumitem}
\usepackage{setspace}
\usepackage[labelfont=bf, singlelinecheck=false, justification=justified]{caption}
\usepackage{wrapfig}
\usepackage{comment}
\usepackage{multicol}
\usepackage[normalem]{ulem}
\usepackage{parskip}  
\usepackage{multirow}
\usepackage[table,xcdraw,dvipsnames]{xcolor}
\usepackage[linewidth=0.5pt,skipabove=10pt,skipbelow=10pt,leftmargin=0pt,rightmargin=0pt,framemethod=TikZ]{mdframed}
\definecolor{myblue}{RGB}{25, 118, 210}
\mdfdefinestyle{bluebox}{backgroundcolor=myblue!10,roundcorner=8pt,linecolor=myblue}
\usepackage{pdfpages}  
\usepackage{soul,color} 

\title{On-sky results from REDWOODS, a platform at Lick/ShaneAO for testing second stage AO technologies}

\author[1]{Benjamin L. Gerard}
\author[1]{Dominic F. Sanchez}
\author[2]{Aditya R. Sengupta}
\author[2]{Chris Ratliff}
\author[3]{Sylvain Cetre}
\author[2]{Elinor Gates}
\author[3]{Daren Dillon}
\author[1]{Bautista Fernandez}
\author[1]{Brian Bauman}
\author[1]{Lisa Poyneer}
\author[1]{Aaron Lemmer}
\author[2]{Rebecca Jensen-Clem}
\author[2]{Phillip Hinz}
\author[2]{Bruce Macintosh}
\affil[1]{Lawrence Livermore National Laboratory}
\affil[2]{University of California Santa Cruz}
\affil[3]{Wakea Consulting}

\authorinfo{Further author information. Send correspondence to Benjamin L. Gerard: \texttt{gerard3@llnl.gov}
}

\begin{document} 
\maketitle

\vspace{3cm}

\begin{abstract}
REDWOODS is a new sub-testbed within the Shane Adaptive Optics (AO) system at Lick observatory, developed to test second stage AO wavefront sensing and control  technologies, including a Self-Coherent Camera with optional broadband Wynne corrector, two different three-sided fully reflective highly broadband pyramid wavefront sensor modes, and multi-wavefront sensor single conjugate AO control. We present project results to date, including on-sky data. 
\end{abstract}


\section{INTRODUCTION}
\label{sec:intro}
The 2020 Decadal Survey of Astronomy and Astrophysics\cite{NASEM2021} prioritized enabling habitable exoplanet imaging by developing emerging technologies in the coming decades for both ground- and space-based observatories. In this proceeding we address this need by presenting in-progress preliminary results from REDWOODS (Real time Exoplanet Direct imaging via Wavefront control Of Optical DefectS), an exoplanet imaging technology demonstrator at Lick Observatory. REDWOODS content was also previously published in 2025\cite{Gerard2025,Hurtado2025}, and we build on these results here. 
\section{HARDWARE OVERVIEW}
\label{sec:overview}
\begin{figure}[!h]
    \centering
    \includegraphics[width=0.75\linewidth]{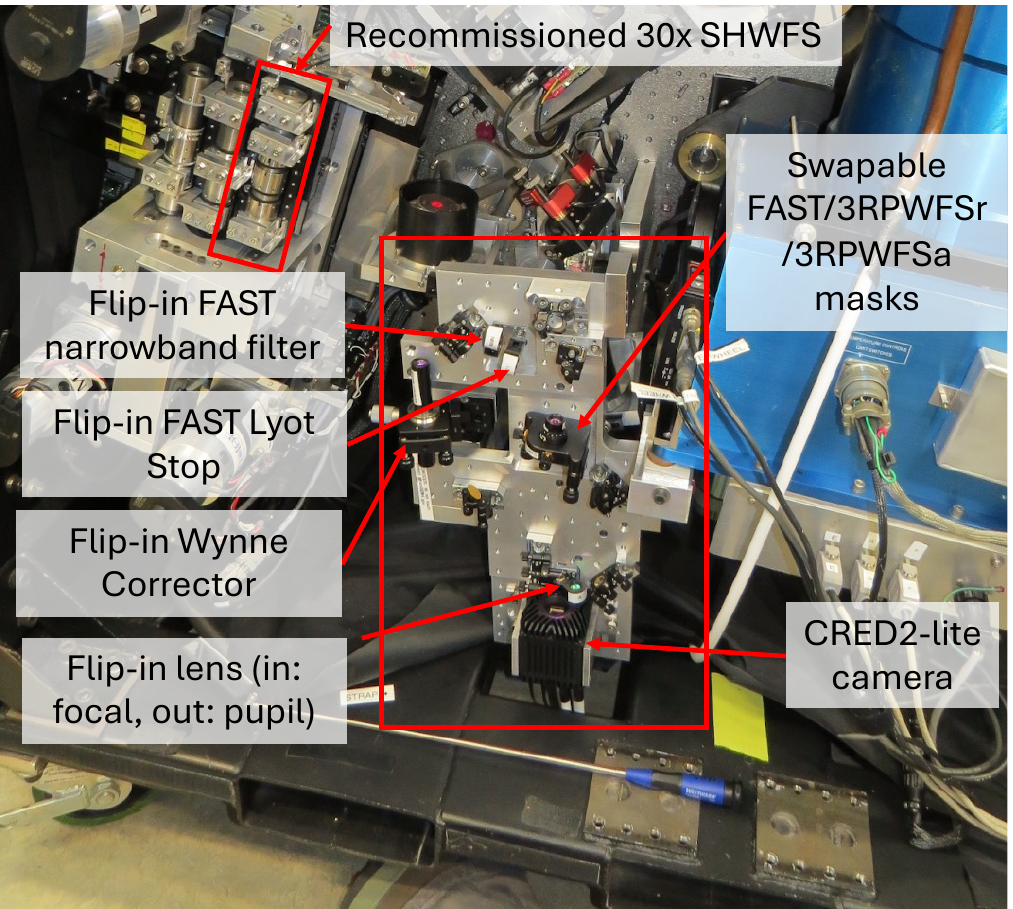}
    \caption{The REDWOODS sub-bench of ShaneAO. All other non-specified optics are OAPs or mirrors. A new RTC server machine, NKT Super COMPACT white light laser are not pictured but also were implemented for this REDWOODS project.}
    \label{fig:redwoods_bench}
\end{figure}

REDWOODS is a sub-bench on the Shane adaptive optics (AO) system\cite{Gavel2016}, shown in Fig. \ref{fig:redwoods_bench}.
REDWOODS implements two modes: (1) a self-coherent camera (SCC)\cite{Baudoz2006}, and (2) a fully-reflective 3-sided Pyramid Wavefront Sensor (3RPWFS), building on previous related 3RPWFS developments at Lick/ShaneAO\cite{Sanchez2024a}. The SCC mode uses the Fast Atmospheric SCC Technique (FAST)\cite{Gerard2018} and baselines a $\Delta\lambda/\lambda_0\sim$2.5\% mode at $\lambda_0=1\mu$m, with an optional Wynne corrector mode with $\Delta\lambda/\lambda_0\sim$21\% mode at $\lambda_0=1.08\mu$m, building off of previous visible Wynne corrector developments and testing\cite{Sanchez2024b}. The REDWOODS Wynne corrector optical design is presented in Ref. \citenum{Sanchez2025}.

The REDWOODS periscope bench was custom-designed to minimize vibrations. Accelerometer measurements with the Shane telescope slewing were taken at the designed sub-bench mounting points before bench installation at ShaneAO and fed into ANSYS finite element analysis (FEA) to predict nm-level vibration stability with the baseline design, well within REDWOODS' requirements to not blur out SCC fringes. Similar FEA also showed negligible gravity-dependent flexure on the ShaneAO+REDWOODS (a Cassegrain instrument port) at the sub-pixel level. Thermal FEA suggested that multi-pixel level drifts over $\sim$hours long timescales were possible due to the CRED2-lite's passive thermo-electric cooling mechanisms and power consumption, but this was considered acceptable as the main goals for REDWOODS are short on-sky real-time AO control demonstrations rather than long exposures.

Most other parts other than the sub-bench were commercial off-the shelf (COTS), including all OAPs, mirrors, the one lens, and every optic-holding mount (all \O 1/2''). The custom Tip/tilt Gaussian (TG) focal plane mask (TG) and 3RPWFS mask was fabricated by Zeiss and validated at 3\% etching depth errors. The Wynne corrector triplets were fabricated by Firebird Optics. Some small custom assembly parts were made in the UCO machine shop. 
\section{INTEGRATION AND TESTING AND INSTALLATION OVERVIEW}
\label{sec:iandt}
\begin{figure}[!h]
    \centering
    \includegraphics[width=0.7\linewidth]{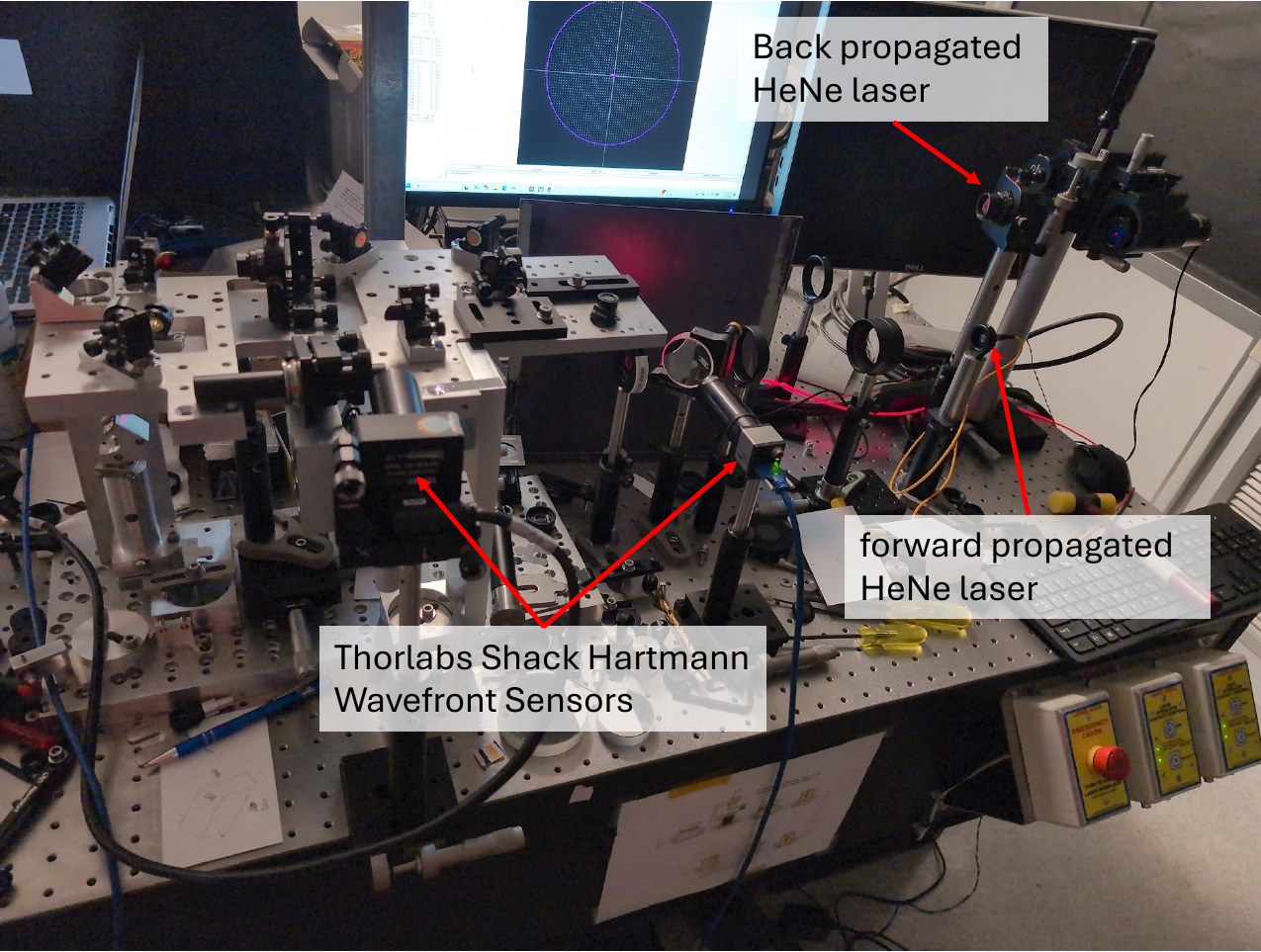}
    \caption{REDWOODS I\&T setup.}
    \label{fig:IandT}
\end{figure}

REDWOODS was designed to pickoff the ShaneAO light with a flip-in mirror downstream of an OAP that would otherwise focus the light to ShARCS, the ShaneAO facility science camera. As such, integration and testing (I\&T) was setup with a matching $\sim$f/30 simulated telescope beam speed to then align REDWOODS. Fig. \ref{fig:IandT} shows the REDWOODS I\&T setup, carried out at UCSC's Laboratory for AO (LAO). 
The bench was designed with positions for targets to align the optical axis to spec with machine precision. We aligned the FPM/3RPWFS mask plane with a knife edge test. OAP clocking was first separately aligned using a SHWFS, and then aligned in-system also with a shear plane and SHWFS. However, the 0.43mm pupil size on the CRED2-lite in pupil imaging mode, which is too small to use a COTS shear plate or SHWFS, made collimation verification challenging to verify. Initially to address this we tried a double-pass approach where the camera plane was replaced by a mirror and a beam cube sent back-propagated light to a SHWFS, which got the system close to alignment, but ultimately we found a single-pass back-propagated approach to work better as shown in Fig. \ref{fig:IandT}. The forward and backward propagated beams were checked for co-alignment through targets and on SHWFSs at various intermediate positions, and then ultimately the forward beam was blocked with a fold to collect only the back-propagated light, which is shown on the SHWFS display in Fig. \ref{fig:IandT} and ultimately resulted in an aligned integrated system wavefront error of 30 nm rms for 12 Zernike modes (piston+tip+tilt removed).
REDWOODS was then transported to the Shane telescope and installed and co-aligned to the existing ShaneAO bench, which is sideways with respect to gravity and thus makes alignment challenging. Anticipating this challenge, before full system installation we initially tested the real-to-CAD model errors with a custom jig to hold 
the first REDWOODS pickoff mirror and position a FLIR camera downstream at the designed focus, which revealed below 1mm discrepancies for the OAP focal length, and optical z and y position, but showed ~6 mm errors for the x position. This informed our full REDWOODS alignment plan, which ultimately used manual nudgers pushing on temporary wings connected to the bench leg mounting points with sufficient adjustability to bring (1) the flip-in REDWOODS mirror into alignment (which had only a $\sim$1mm x and y displacement margin for error due to tight space constraints, and similar z tolerance for depth of focus. 

Separately, the original ShaneAO ``30x'' mode (which is actually 29 subapertures across the on-sky pupil) was reinstalled as the third configurable mode of the ShaneAO SHWFS path. This mode was previously decommissioned and replaced by a 3RPWFS.\cite{Sanchez2024a} However, it was of interest for this project to re-install this 30x mode to reduce fitting error and increase Strehl fed to REDWOODS. The parts for this mode, all contained within a single barrel to ensure displacement co-alignment, include a collimating lens, lenslet array, and 1:1 magnification relay designed to produce $\sim$0.6 pixels per $\lambda_0/d$ and 5 pixel subaperture pitch. Initial system alignment in the LAO similarly setup a telescope simulator to match the design input f number, and imaged the visible SHWFS spots onto a FLIR camera that appeared to produce the right spot separation at the few $\mu$m level of designed specifications, which was then installed in ShaneAO in January 2026. However, subsequent processing of SHWFS images via the demodulation technique\cite{Ribak2006} indicated the subaperture pitch was around 4.9 pixels instead of the designed 5.0. On July 22, 2026, we used this same technique to dial in the subaperture pitch to 4.99 in both x and y and lenslet-to-WFS camera pixel clocking offset to 0.1 degrees, which will likely make a difference in future on-sky performance and may have been at least in-part why performance has been sub-optimal in runs so far (see Appendix \ref{sec:slopes} for related discussion).
\section{REAL-TIME SOFTWARE OVERVIEW}
\label{sec:software}
\begin{figure}[!h]
    \centering
    \includegraphics[width=0.6\linewidth]{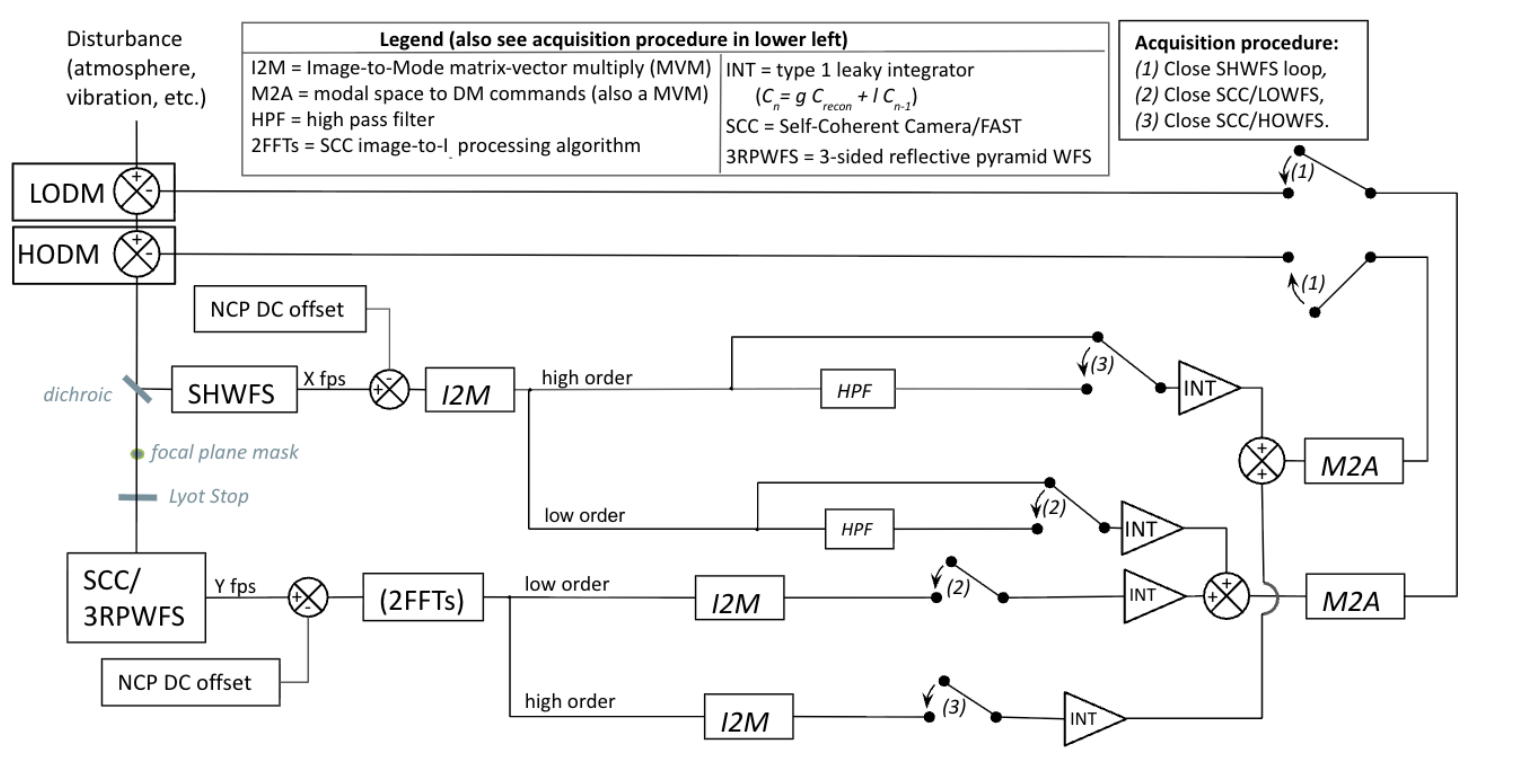}
    \includegraphics[width=0.39\linewidth]{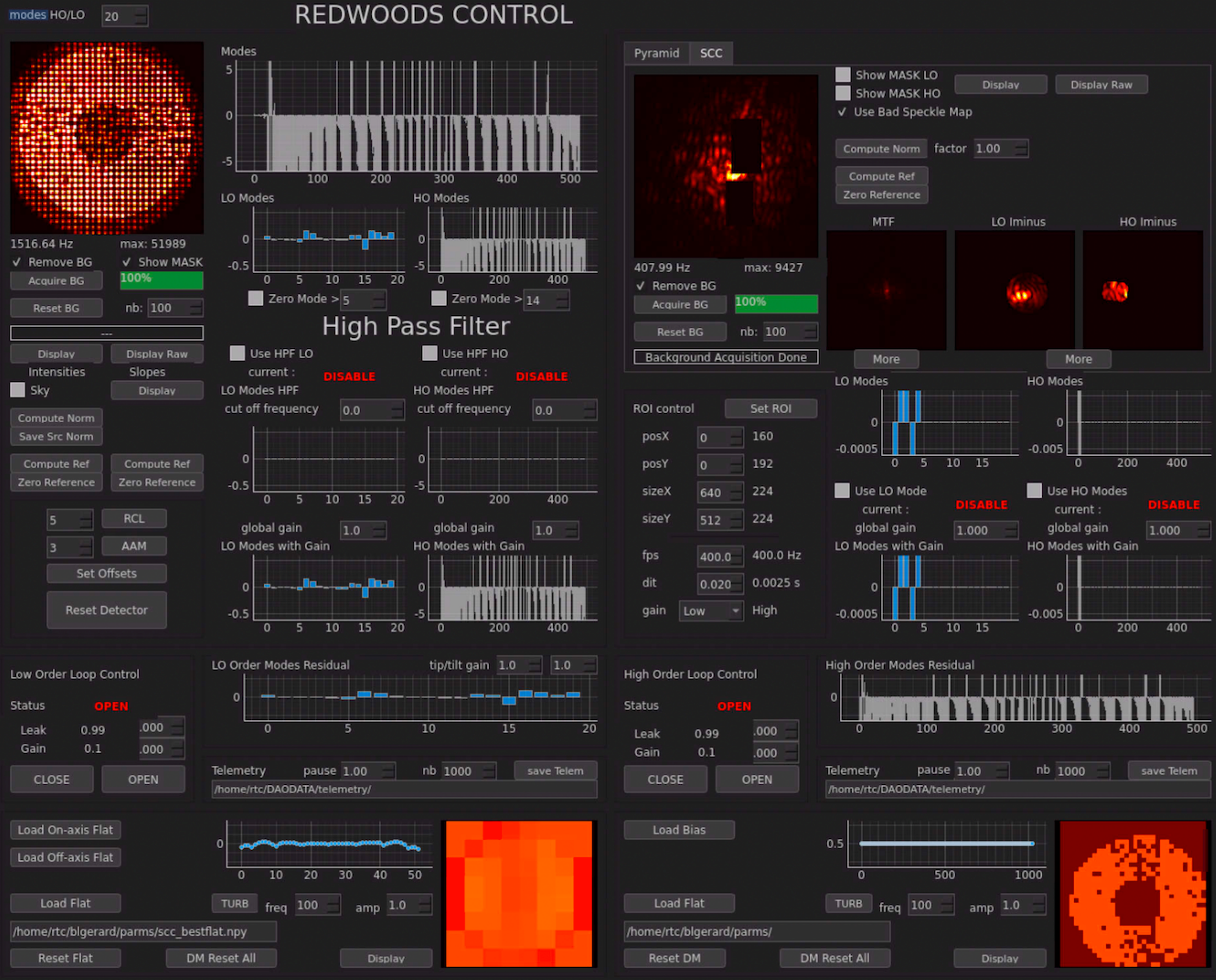}
    \caption{Left: high-level multi-WFS control architecture diagram, from Ref. \citenum{Gerard2025}. Right: Actual used multi-WFS control software GUI that realizes left diagram.}
    \label{fig:control_gui}
\end{figure}

REDWOODS aims to implement multi-WFS single conjugate AO control, where two or more WFSs both looking at the same star control one or more common path DM. Ref. \citenum{Sengupta2026} analyzes this generic AO control problem, showing that temporal non-common path aberrations (NCPAs; e.g., different air blowing in the two WFS paths) must be below 50 nm rms to implement classical double integral control. Above this 50 nm rms ``hairdryer limit,'' additional solutions such as high past filtering are needed to minimize inter-arm temporal NCPA transfer. For REDWOODS, the ShaneAO 30x SHWFS and FAST/3RPWFS are the two WFSs, and the ShaneAO 32x32 tweeter (hereafter HODM) and 52 actuator woofer (hereafter LODM) are the common path DMs. Because both WFSs must share a common modal structure to control the same common path DMs, it was decided early on in the REDWOODS project that we would re-write our own SHWFS software and not use the facility ShaneAO software, and additionally, to not interrupt facility AO operations, such code would be developed on a separate RTC machine, which was purchased initially to support REDWOODS but ultimately to update ShaneAO software into a more modern software environment consistent with more recent open source AO real-time code development capabilities. Details of this new ShaneAO/REDWOODS RTC, called ``tooreal'' (ultimately replacing the ``real'' RTC) will be described in another paper. A block diagram of the the REDWOODS multi-WFS control architecture (initially from Ref. \citenum{Gerard2025}) and a snapshot of the GUI implemented to realize this for the REDWOODS project are shown in Fig. \ref{fig:control_gui}
REDWOODS implements a matrix vector multiple (MVM) reconstructor approach for both WFSs, where pixel-level information for each WFS (on each WFS's internal clock that does not require synchronization or integer-multiple frame rates between WFSs) can be separately MVM'd into modal space (``I2M'' in Fig. \ref{fig:control_gui}), which when shared by both WFSs can then be added and then MVM'd again into DM command space (``M2A'' in Fig. \ref{fig:control_gui}). This architecture was realized using DAO RTC pipeline,\footnote{\url{https://www.eso.org/sci/meetings/2023/RTC4AO/02_08_cetre_s.pdf}} which implements backend C-based shared memory processes and enables seamless interfacing of high-level and soft real-time code in Python, as has been demonstrated in the past on related code for FAST\cite{Gerard2022}. A GUI version displaying capabilities and functionalities of all the back-end real-time processes is in Fig. \ref{fig:control_gui}. We used this pipeline with the new RTC machine to benchmark the entire pipeline in reduced intensity SHWFS 30x mode (see \S\ref{sec:calibration} for further discussion) to benchmark $\sim$250 $\mu$s computational latency, clearly justifying that the system is fast enough for real-time AO control.
\section{CALIBRATION}
\label{sec:calibration}
\begin{figure}[!h]
    \centering
    \includegraphics[width=0.54\linewidth]{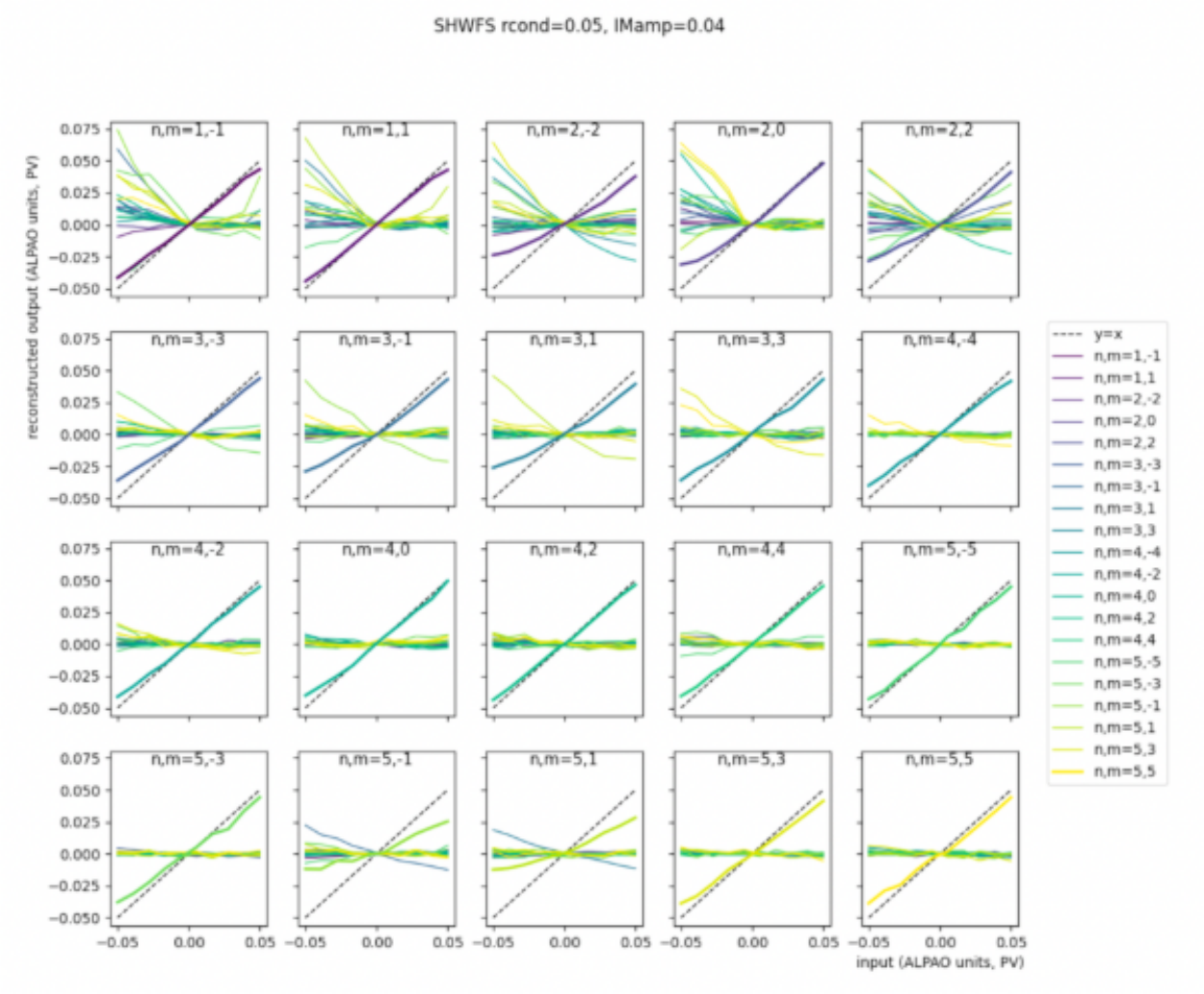}
    \includegraphics[width=0.45\linewidth]{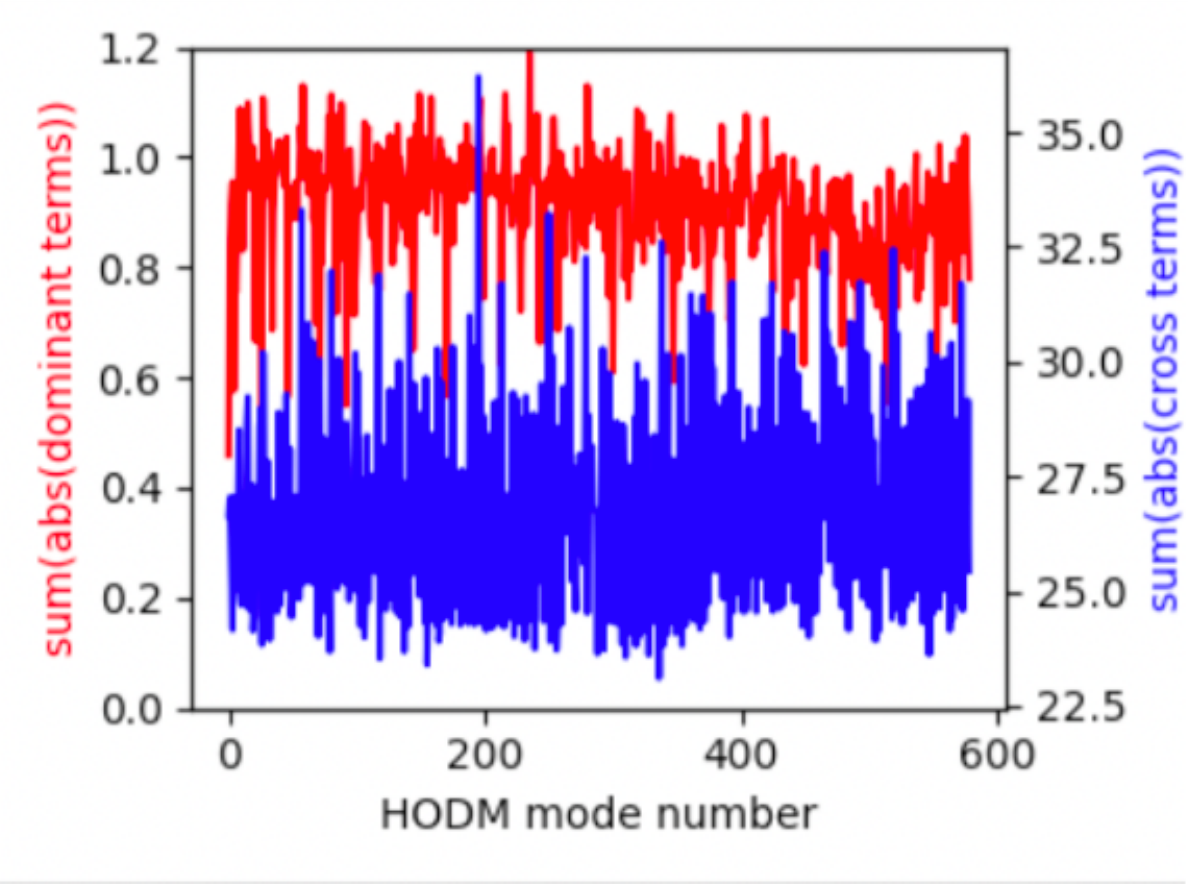}
    \caption{LODM (left) and HODM (right) non-linearities from SHWFS 30x mode (before re-alignment) from internal source reduced intensity operations.}
    \label{fig:linearities}
\end{figure}

\begin{figure}[!h]
    \centering
    \includegraphics[width=0.95\linewidth]{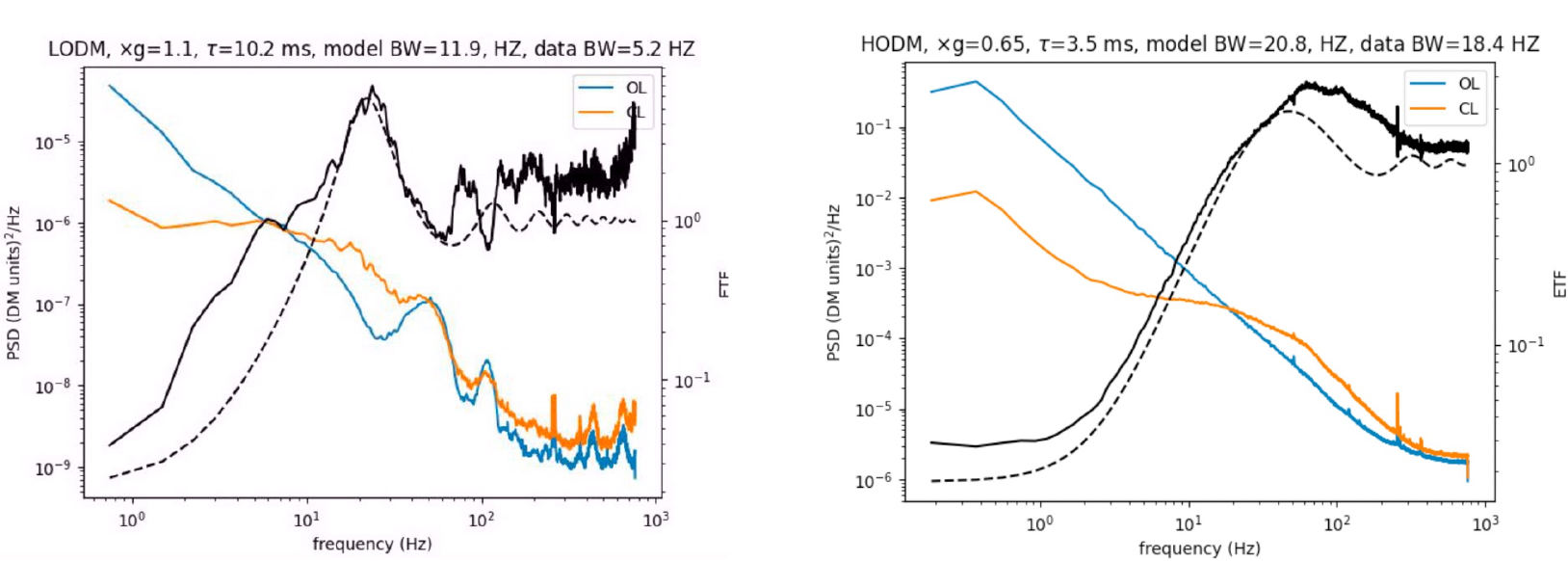}
    \caption{HODM and LODM SHWFS 30x mode (before re-alignment) ETFs from internal source reduced intensity operations.}
    \label{fig:etfs}
\end{figure}

Motivated by Ref. \citenum{Gerard2025}, we initially baselined a reduced intensity 30x SHWFS mode, meaning no slopes were calculated and instead SHWFS image pixels, normalized by the total image flux and then temporally medianed over a user-defined number of frames, are directly MVM'd into modal space. As just discussed at the end of \S\ref{sec:software}, despite this approach using 12.5$\times$ as many pixels, we benchmarked a sufficiently low computational latency to motivate using this technique. To build the I2M, we performed brute force optimization to toggle various hyper-parameters to ultimately produce acceptable linearity metric, shown in Fig. \ref{fig:linearities}.
Linearity curves in Fig. \ref{fig:linearities} are plotted for individual LODM Zernike modes but a proxy metric is instead shown for HODM modes as there would otherwise be too many modes to plot. The equivalent linearity data is taken for every HODM mode (which is a zonal poke with the 20 LODM Zernike modes least-suqares deprojected to form an orthogonal basis), where then the red curve in Fig. \ref{fig:linearities} shows the sum of the absolute value of the linearity data of the given probed mode and the blue curve shows the equivalent sum for all the cross terms for each given mode. When the red and blue curves each have a relatively low standard deviation, and spot checks of individual HODM mode linearity curves look satisfactory, it can be assumed that overall HODM calibration is satisfactory. After HODM and LODM linearities are validated, we used respective DM-injected turbulence of each mode with a -2 temporal power law and amplitude and ensuring that the amplitudes were less than 50\% non-linear (reconstructed amplitude relative to the y=x line) and the respective I2Ms to computed modal coefficients for open-loop and closed loop temporal power spectral density (PSD) plots and ultimately produce SHWFS error transfer functions (ETFs) as shwon in Fig. \ref{fig:etfs}.
The PSDs and ETFs in Fig. \ref{fig:etfs} are an average of all modes in each group for a 9000 frame dataset of open and separately closed loop operations, which then allows modal averaging to further reduce noise before ETF fits can be comared to data. However, as Fig. \ref{fig:etfs} shows the fits do match the data well and thus cannot be used for optical gain compensation as was intended initially and as was done in Ref. \citenum{Gerard2025} (but again see further discussion in \S\ref{sec:slopes}), but regardless the measured 0 dB bandwidths are at least consistent with equivalent facility ShaneAO bandwidth measurements,\cite{Rudy2017} suggesting that on-sky bandwidth error should be similar to facility ShaneAO software. Unfortunately, the $\sim$10 ms LODM system latency suggested by the ETF fit is also consistent with Ref. \citenum{Rudy2017}; the LODM is connected to another computer that is connected to the main RTC via ethernet, thus causing this high latency mainly due to network traffic limits.

\begin{figure}[!h]
    \centering
    \includegraphics[width=0.99\linewidth]{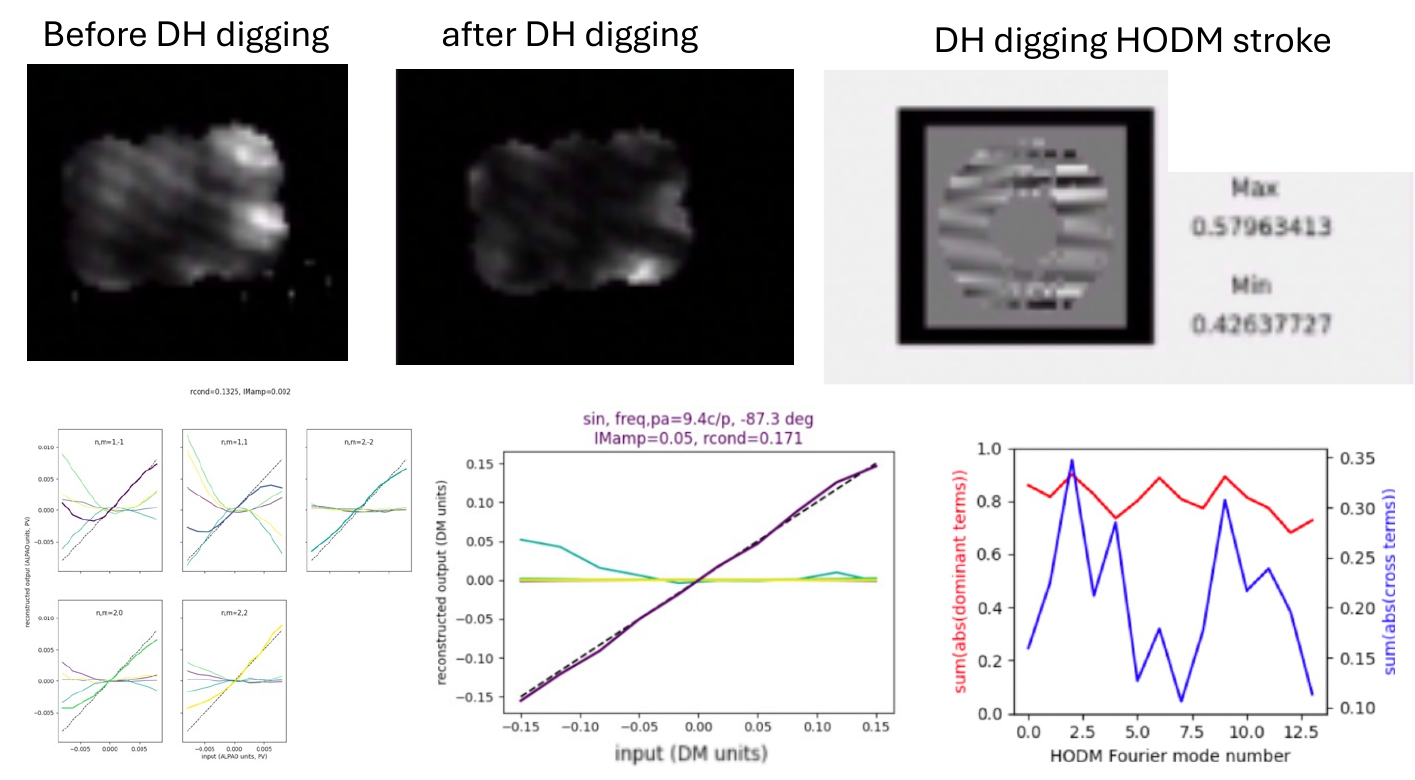}
    \caption{FAST calibration on internal source in narrowband mode. Upper left and middle: before and after, respectively, dark hole (DH) digging, both on the same spatial and contrast scale. Upper right: fractional HODM stroke used for dark hole ($\sim$15\%). Lower left: FAST LODM linearity. Lower middle and right: FAST HODM linearity}
    \label{fig:scc_cal}
\end{figure}

A similar MVM pipeline and linearity calibration is implemented for the narrowband SCC mode (PWFS mode linearity has thus far been validated for a MVM that uses the full 640x512 CRED2-lite array size and is thus not yet ready for real-time control, with modifications making it ready to be presented in a future paper), shown in Fig. \ref{fig:scc_cal}.
Unfortunately, a large number of HODM bad actuators, dust and amplitude errors, and perhaps other factors (see \S\ref{sec:challenges}) limited dark hole digging to a relatively small 2$\times$3 $\lambda/D$ area. Regardless, Fig. \ref{fig:scc_cal} shows that LODM linearity (which uses the SCC image core and not the dark hole pixels\cite{Gerard2025}) and HODM linearity indicate they are ready for on-sky closed-loop real-time control if the SHWFS loop can bring residual AO wavefront errors within the SCC capture range.
\section{PRELIMINARY IN-PROGRESS ON-SKY RESULTS}
\label{sec:on-sky}
\begin{figure}[!h]
    \centering
    \includegraphics[width=0.99\linewidth]{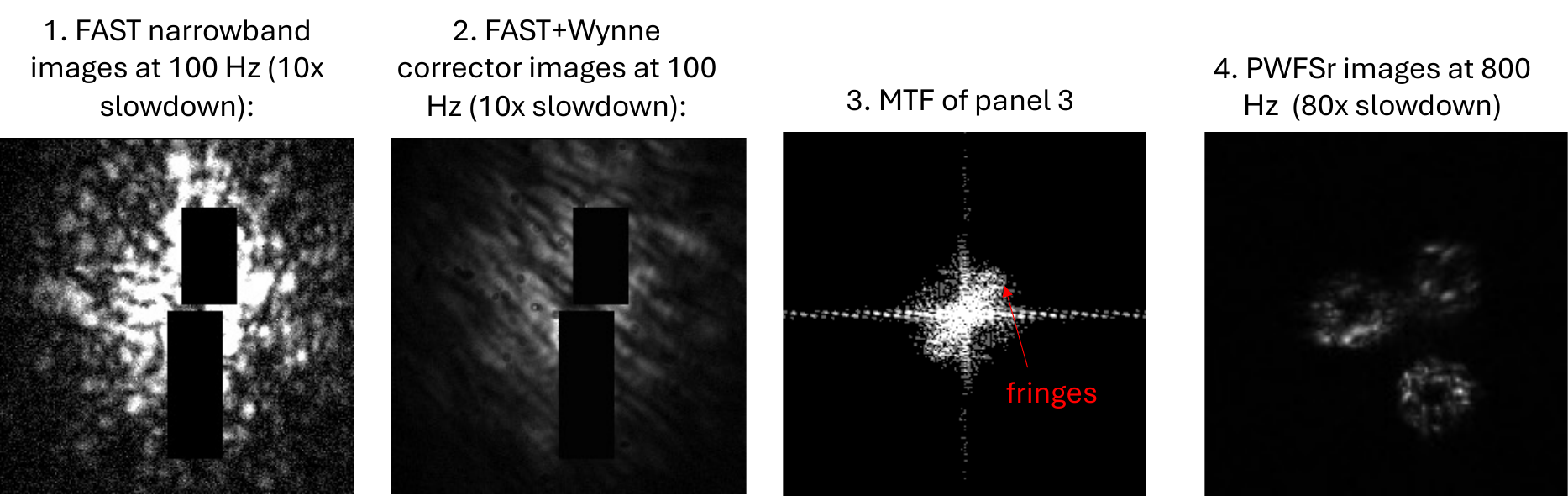}
    \caption{Preliminary on-sky REDWOODS images from June 29 on-sky observations. See online presentation for movies corresponding to these images.}
    \label{fig:onsky}
\end{figure}

As of this writing in late July 2026, project timeline and weather have enabled two nights of on-sky observations, during which clouds were often present and seeing fluctuated between 1-3" (via ancillary measurements made by other Lick observatory instruments, so such numbers should be considered qualitatively). The 30x SHWFS loops were closed on-sky in reduced intensity mode, and REDWOODS image data cubes in various modes were saved and shown in Fig. \ref{fig:onsky}.
It was clear when loops were closed that the PWFS alignment ``snapped'' into place that second stage PWFS loops (if ready, which they are not yet) could potentially close, but in SCC mode it appears that the AO residual wavefronts were not diffraction-limited, with loops closing nicely to a defined aligned reference position on the internal source with the telescope tracking (even for strong DM-injected turbulence), but such loops not closing nearly as well on-sky just after that. We believe this was due to an improper normalization technique in reduced intensity mode and will also be further improved with a more conventional slopes pipeline instead of reduced intensity, which we discuss further in \S\ref{sec:slopes}. 

Various bright stars were observed to ultimately determine that $m_J\lesssim0$ magnitude stars are needed to detect $\sim$tens of photons per AO residual speckle (consistent with past related simulations, e.g., \citenum{Gerard2018}) such that with $\gtrsim$10\% fringe visibilities (which we have corroborated to be true on the internal source) fringes can be detected on such speckles in narrowband mode. However, in broadband mode with our Wynne corrector in the beam, Fig. \ref{fig:onsky} shows that we have clearly detected fringes on-sky at 100 Hz, enabled by the higher throughput opened up by the 8.4$\times$ larger spectral bandpass of this mode. Although fringes are detected, speckles are significantly smeared out and likely their strehl is too low to motivate future exoplanet imaging science applications. We are currently working on calibrating a tradeoff mode where the Wynne corrector off-axis displacement is halfway in between the Lyot stop pinhole and pupil aperture to balance the boost of fringe visibility vs. loss of Strehl. 
\section{CONCLUSION}
\label{sec:conclusion}
REDWOODS is a second stage adaptive optics technology demonstrator for exoplanet imaging capabilities, deploying multiple new diffraction-limited WFS capabilities aimed to boost Strehl and/or contrast to enable detection of higher contrast and/or closer-in exoplanets. Thus far the project has detected on-sky fringes in FAST/SCC broadband mode at 100 Hz, which suggest at minimum that such fringes can be used for coherent differential imaging post-processing to overcome the fundamental gap relative to the photon noise limit that is present when taking longer exposures that average AO residual turbulence.\citenum{Gerard2018} We are part way through the project's on-sky validation phase and have closed the reduced intensity SHWFS loops with some gains but not yet diffraction-limited performance. These on-sky results will be further improved during upcoming observing runs in August and September 2026.
\appendix
\section{SLOPES VS. REDUCED INTENSITY SHWFS}
\label{sec:slopes}
We used our internal AO simulator code, built up and validated against decades of AO testbed and on-sky development (same code used in Ref. \citenum{Sengupta2026}; see this reference for further descriptions), to determine that a reduced intensity SHWFS can close the loop on full-strength turbulence and produce diffraction-limited PSFs, but only when a specific type of normalization is implemented. Because the flux per subaperture will likely be different between on-sky and internal source, we divided the image pixels by the sum of the image, medianed over 1000 frames, but for on-sky performance, despite implementing an equivalent normalization, we only found best performance on stars where the subaperture flux was around the same as with the internal source. The reason for this is that ``numerator'' subaperture pixels, e.g., in ADU, have a different response for a given absolute wavefront error depending on the brightness of the source, even with the above-described normalization. For an internal source SHWFS image that is normalized by $I_\text{source}/\Sigma I_\mathrm{source}$, where $\Sigma$ represents the median averaged total flux of the image, where such normalization is then used to generate a command matrix that will by MVM'd by on-sky SHWFS pixels with an equivalent normalization, the needed on-sky image must be normalized as $I_\text{sky}\frac{\Sigma I_\mathrm{source}}{(\Sigma I_\mathrm{sky})^2}$ to produce diffraction-limited performance, and for our on-sky implementation thus far we have only implemented a $I_\text{sky}/\Sigma I_\mathrm{sky}$-based normalization. We have made this change in our software and expect improved performance in reduced intensity SHWFS mode for the next on-sky run.

\begin{figure}[!h]
    \centering
    \includegraphics[width=0.49\linewidth]{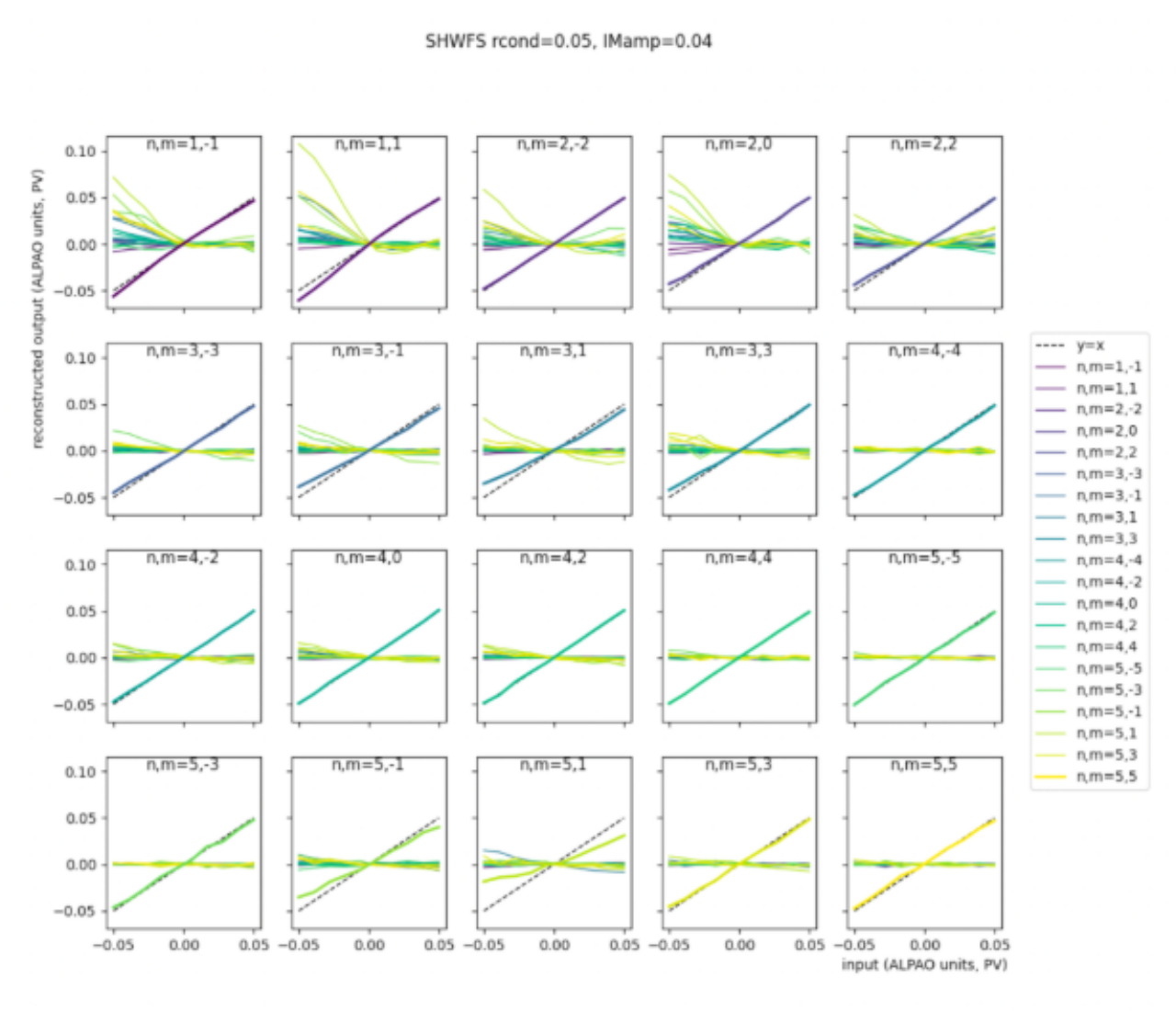}
    \includegraphics[width=0.49\linewidth]{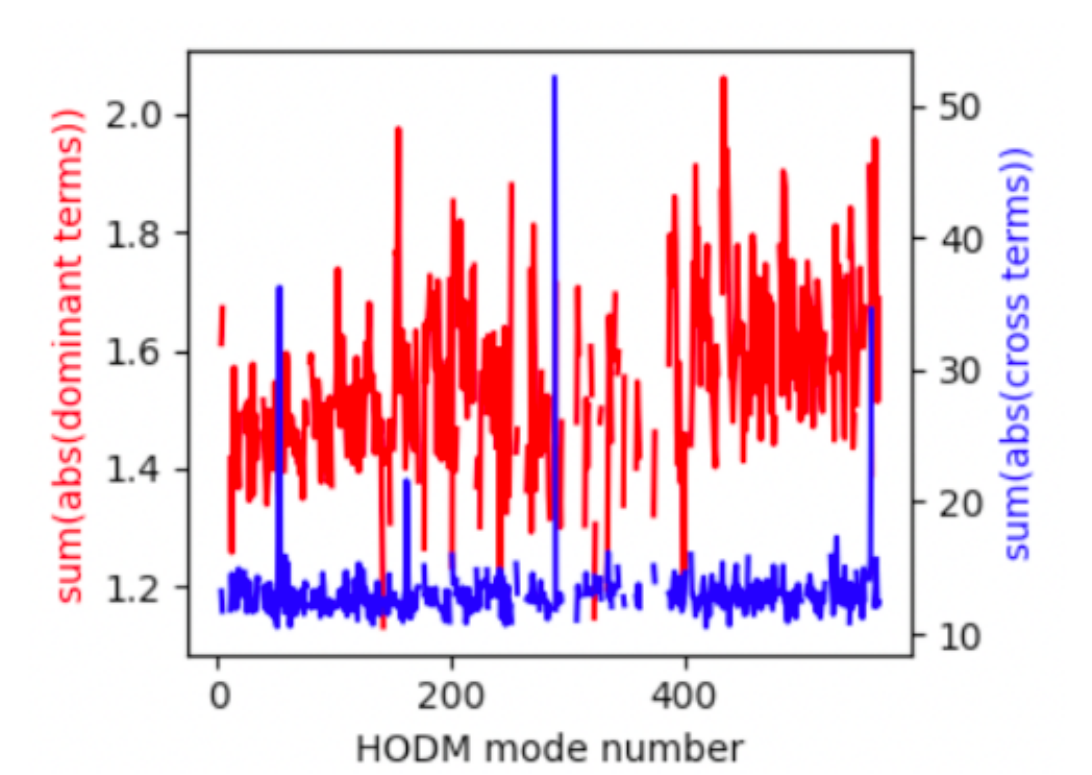}
    \caption{Slopes LODM (left) and HODM (right) linearities }
    \label{fig:slopes_linearity}
\end{figure}

\begin{figure}[!h]
    \centering
    \includegraphics[width=0.95\linewidth]{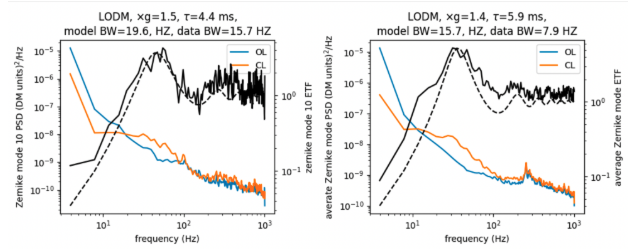}
    \includegraphics[width=0.95\linewidth]{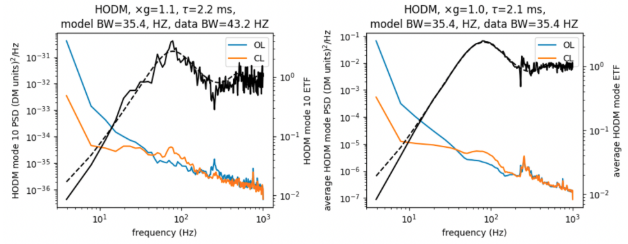}
    \caption{Internal source SHWFS ETFs using slopes instead of reduced intensity, showing low order modes (top row) and high order modes (bottom row) for a single mode, mode 10 (left column) and an average of all modes (right column). Bandwidths are clearly higher and fits are better than reduced intensity, motivating future use of slopes.}
    \label{fig:SlopesETFs}
\end{figure}

We have also realigned the 30x SHWFS mode using a SHWFS demodulation technique as mentioned in \S\ref{sec:iandt} to quantitatively fine tune the SHWFS subaperture pitch to 4.99 pixels, very close to 5.0 and now enabling a standard center of gravity slopes processing pipeline mode. With that mode, the LODM and HODM linearities are in Fig. \ref{fig:slopes_linearity}. 
Comparing Fig. \ref{fig:slopes_linearity} to Fig. \ref{fig:linearities} suggests that slopes mode will have better performance and reduce optical gain effects. Further corroborating the liklihood of increased performance with SHWFS slopes vs. reduced intensity, internal source slopes ETFs are shown in Fig. \ref{fig:SlopesETFs}.
\section{Technical Challenges}
\label{sec:challenges}
\begin{figure}[!h]
    \centering
    \includegraphics[width=0.5\linewidth]{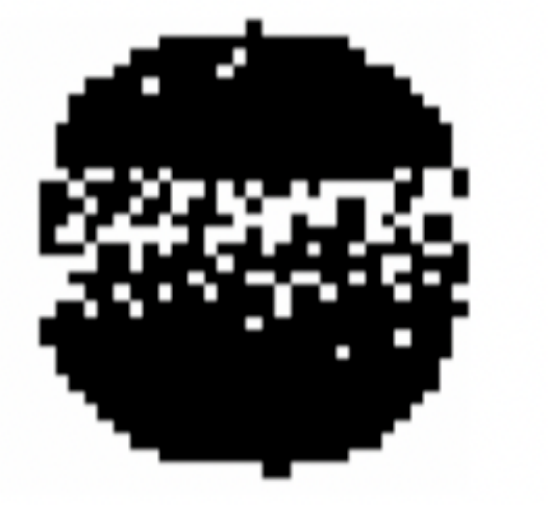}
    \caption{Map of bad HODM actuators. Of the $\sim$804 controllable actuators within the inscribed circular aperture, $\sim$200 are stuck at 0.}
    \label{fig:badactuators}
\end{figure}

Two key technical challenges are worth mentioning here, particularly in light of ShaneAO's increased use and interest for future technology development projects. First, Our re-installation of the 30x mode revealed $\sim$200 bad HODM actuators, visualized in Fig. \ref{fig:badactuators}.
It is not surprising that this has happened to a relatively old AO system which implements one of the earliest generation 1k BMC MEMS DMs that does not have a sealed window and is now used for facility ShaneAO operations with humidity constraints that would otherwise prevent bad actuators from being generated. Using a newer BMC HODM with a chemcially sealed window with purged dry air is likely to address this problem. Regardless, until such an upgrade happens, one can assume that fitting error is $\sim$25\% worse than an equivalent 100\% yeild HODM, at least when controlling modes out to the DM Nyquist limit with the 30x mode. 

Secondly, the ShaneAO optics have not been replaced or recoated since their installation, and some surfaces generate significant dust/ringing in images. This makes internal light source placement and dark hole digging challenging in a ``dust mine field''-like environment. The Shane telescope does have dome flat capabilities which we will try to use to calibrate out speckles in the our next observing run, but for daytime off-telescope work (which constitutes the majority of software work done for this project), a flat field source option could be of beneifit to future shaneAO projects in adition to re-coating at least a key ``key offender'' optics.
\vspace{2cm}
\acknowledgments 
 
This work was performed under the auspices of the U.S. Department of Energy by Lawrence Livermore National Laboratory under Contract DE-AC52-07NA27344. This document number is LLNL-PROC-2022303. This work was jointly funded by NSF ATI award \# 2520860, LLNL LDRD tracking number 24-LW-002, and UC Observatories/LLNL SPP contract L24118.

\vspace{2cm}

%
\bibliography{report} 
\bibliographystyle{spiebib} 

\end{document}